\documentclass[12pt]{iopart}
\usepackage{iopams}

\usepackage{xcolor}
\usepackage{graphicx}

\expandafter\let\csname equation*\endcsname\relax

\expandafter\let\csname endequation*\endcsname\relax

\usepackage{amsmath}
\usepackage{mathrsfs}
\usepackage[T1]{fontenc}
\newcommand{\thorn}{\mbox{\th}}
\usepackage{float}
\usepackage{subfig}
\usepackage{tikz}
\usetikzlibrary{decorations.pathmorphing}
\usetikzlibrary{arrows.meta}

\usepackage[pdfencoding=auto, psdextra,pdftex,pdfpagelabels]{hyperref}

\usepackage[
            style=phys,
            sorting=nyt,
            articletitle=true,
            biblabel=brackets,
            chaptertitle=true,
            pageranges=true,
            hyperref=true,
            maxcitenames=3,
            maxbibnames=4,
            url=true,
            autocite=superscript
            ] {biblatex}

\usepackage[final]{showkeys}

\newcommand{\eps}{\varepsilon}
\newcommand{\del}{\partial}

\newcommand*{\ul}[1]{\underline{#1}}

\newcommand*{\bftm}{\widetilde{\mathbf{m}}}
\newcommand*{\bftmb}{\overline{\widetilde{\mathbf{m}}}}

\newcommand*{\tdelta}{\widetilde{\delta}}

\newcommand*{\teth}{\widetilde{\eth}}

\newcommand*{\hM}{\hat M}

\newcommand*{\cQ}{\mathcal{Q}}
\newcommand*{\cR}{\mathcal{R}}

\newcommand*{\CC}{\mathbb{C}}

\newcommand*{\tM}{\widetilde M}
\newcommand*{\tg}{\widetilde g}
\newcommand*{\tR}{\widetilde R}

\newcommand*{\tpsi}{\widetilde \psi}
\newcommand*{\hnabla}{\widehat\nabla}

\newcommand*{\hdal}{{\Box\kern-7.5pt\wedge}}

\newcommand*{\scri}{\mathscr{I}}
\newcommand*{\const}{\mathit{const.}}
\newcommand*{\ii}{\mathrm{i}}

\begin{document}
\title[The non-linear perturbation of a black hole by gravitational waves]{The non-linear perturbation of a black hole by gravitational waves. I. The Bondi-Sachs mass loss}

\author{J Frauendiener$^1$ and C Stevens$^1$}

\address{${}^1$Department of Mathematics and Statistics, University of Otago, Dunedin 9016, New Zealand}

\ead{joergf@maths.otago.ac.nz, cstevens@maths.otago.ac.nz}

\begin{abstract}
The excitation of a black hole by infalling matter or radiation has been studied for a long time, mostly in linear perturbation theory. In this paper we study numerically the response of a Schwarzschild black hole to an incoming gravitational wave pulse. We present a numerically well-posed initial boundary value problem for the generalized conformal field equations in which a wave profile for the ingoing wave is specified at an outer time-like boundary which then hits an initially static and spherically symmetric black hole. The non-linear interaction of the black hole with the gravitational wave leads to scattered radiation moving back out. The clean separation between initial state and incoming radiation makes this setup ideal to study scattering problems. The use of the conformal field equations allows us to trace the response of the black hole to null infinity where we can read off the scattered gravitational waves and compute the Bondi-Sachs mass and the gravitational flux through $\scri$. In this way we check the Bondi-Sachs mass loss formula directly on null infinity. We also comment on comparisons with quasinormal modes. 
\end{abstract}
\maketitle

\section{Introduction}

With the work by Bondi~\cite{Bondi:1962}, Sachs~\cite{Sachs:1962a}, Newman and Penrose~\cite{Newman:1962,Newman:1962b,Penrose:1963} in the 1960's it has become clear that Einstein's theory admits gravitational waves. It has also become apparent that, just as in Maxwell's theory, the notion of radiation is well defined only at infinitely large distances from a source. Penrose's proposal for the geometric treatment of asymptotically flat space-times~\cite{Penrose:1965} demonstrated that the notion of infinity can be precisely captured by the idea of conformal compactification which stipulates that asymptotically flat space-times are those for which one can conformally rescale the metric in such a way that one can attach a conformal boundary, commonly called null infinity and denoted by $\scri$ (for more details see~\cite{Frauendiener:2004a,Geroch:1977,ValienteKroon:2016}). The mass loss formula proved by Bondi et.al.~\cite{Bondi:1962,Sachs:1962a}, Newman and Penrose~\cite{Newman:1962,Penrose:1963} is a balance law between asymptotic quantities, i.e., it holds at null infinity $\scri$, relating the decrease of energy-momentum contained inside the space-time at a certain instant over time to the gravitational flux radiated away from the source. 

In the new era of gravitational physics which started with the detection of gravitational waves~\cite{Abbott:2016} the fact that wave forms can be computed accurately only at infinity poses a problem, at least in principle. Current commonly employed frameworks for numerically solving the Einstein equations  operate on a finite part of the space-time and thus approximation and limiting procedures must be used (see \cite{Bishop:2016} for an overview.) Nevertheless, these methods have been very successful in practice, in particular for calculating outgoing gravitational waveforms from compact binaries for comparison with observational data from LIGO.

There do exist methods by which the asymptotic quantities can be computed directly without the need for approximation or limiting procedures. They all make use, to varying degrees, of compactification. One of the earliest methods~\cite{Gomez:1992,Gomez:1992a,Bishop:2016} is based on characteristic hypersurfaces which are compactified in order to bring the points at infinity to a finite location. Another method, recently increasingly popular, is based on the hyperboloidal initial value problem in which the space-time is foliated by space-like hypersurfaces which extend to null infinity. These hyperboloidal hypersurfaces are spatially compactified~\cite{Rinne:2010}. The method of Cauchy-Characteristic matching~\cite{Bishop:1993} tries to combine the characteristic method with a Cauchy evolution of the Einstein equations with a matching procedure across a time-like interface. The first two methods have been implemented successfully but have certain drawbacks. The characteristic approach suffers from the fact that the null hypersurfaces may develop caustics in strong fields while the equations used in the hyperboloidal method are singular on~$\scri$.



Friedrich's conformal field equations \cite{Friedrich:1981,Friedrich:1981a,Friedrich:1995} are a system of equations which generalise Einstein's equations by implementing Penrose's proposal of conformal compactification~\cite{Penrose:1965}. In this way, the conformal field equations address an asymptotically simple space-time including its conformal boundary (and beyond).
Recently, an Initial Boundary Value Problem (IBVP) framework for the Generalized Conformal Field Equations (GCFE) \cite{Friedrich:1995} was presented and verified to be well-posed (at least numerically) for a variety of different initial and boundary conditions \cite{Beyer:2017}. In particular, $\scri^+$ was successfully incorporated in the computational domain for the numerical evolution of a Schwarzschild space-time perturbed by gravitational waves. To do so, the equations  were written completely in the space-spinor formalism \cite{Sommers:1980} and Newman and Penrose's $\eth$-calculus \cite{Penrose:1984a,Penrose:1986} for spin-weighted spherical harmonics on the sphere was employed to allow for fast and accurate pseudo-spectral methods to be utilized \cite{Huffenberger:2010}. A method for prescribing boundary conditions was put forward which guarantees constraint propagation and was numerically verified. It was shown that in axisymmetry and using a simple ${}_2Y_{20}$ mode for the ingoing gravitational radiation, the numerical evolution was stable up to and beyond $\scri^+$.

Since the computational domain contains at least a portion of the conformal boundary, one can compute the global energy-momentum quantities there which is what we will focus on in this work. Eventually, each time-slice of this numerical evolution will intersect null infinity in a \emph{cut}.  On each cut, one can calculate the asymptotic quantities such as the components of the Bondi-Sachs momentum 4-covector. However there is a major snag: In the literature (e.g. \cite{Penrose:1986}) judicial choices of the coordinates, frame and conformal factor are made to simplify the mathematical analysis as much as possible. These choices, which are generally made based on geometric considerations, are in general inconsistent with what is required to implement a stable numerical scheme. So then, how does one who has run a numerical evolution that includes $\scri$ in the computational domain actually compute the correct quantities? In a recent paper~\cite{Frauendiener:2021a} we discuss this question and derive expressions for the Bondi-Sachs energy-momentum and the gravitational flux which are valid on arbitrary cuts and for any choice of coordinates, frame and conformal factor.

The aim of this paper is to apply these formulae in a concrete setting and to numerically test them for non-linear gravitational perturbations of the Schwarzschild space-time using the IBVP framework for the GCFE as the numerical evolution scheme. Since this scenario was described in detail in~\cite{Beyer:2017} we refer the reader there for thorough checks of correctness, such as constraint convergence tests etc.

The layout of the paper is as follows: Section \ref{sec:OverviewOfIBVP} provides a summary the IBVP framework for the GCFE whilst in Section \ref{sec:IDandBD} we describe the specifics of the system to be evolved. Section \ref{sec:Bondi4momentum} describes how we make use of the formula for the Bondi components given in \cite{Frauendiener:2021a} and Section \ref{sec:Results} presents the numerical results. The paper is concluded with a brief discussion and a summary in Section \ref{sec:Summary}. We use the conventions of \cite{Beyer:2017} throughout. 

\section{Overview of the GCFE and the IBVP framework}
\label{sec:OverviewOfIBVP}

Here we briefly summarize the most relevant aspects of the GCFE and the IBVP framework for them and point the reader to \cite{Friedrich:2002} and \cite{Beyer:2017} respectively for a more comprehensive review.

Given a space-time $(\tM,\tg)$, thought of as the ``physical'' space-time, Einstein's vacuum field equation with vanishing cosmological constant is given by $\tR_{ab} = 0$. We are interested in the properties of solutions of this equation ``at infinity''. In order to access the points at infinity one introduces an appropriate conformal factor $\Theta$ and defines the ``conformal'' metric as $g=\Theta^2\tg$ and the conformal boundary is then defined as the set of points for which $\Theta=0$ and $\mathrm{d}\Theta\ne0$. This condition defines a regular hypersurface $\scri$ in the conformal space-time  $(M,g_{ab})$ which can be shown to be null in the case of vanishing cosmological constant.

The metric conformal field equations are a regular extension of the vacuum equations incorporating the conformal rescaling of the physical metric. They express the physical content of the Einstein equations entirely in terms of the geometry on the conformal manifold. The relevant quantities are $\{g_{ab},\Theta,s,P_{ab},K_{abc}{}^d\}$ where $s := \frac14\nabla^c\nabla_c\Theta - \frac{1}{24}R\Theta$, $P_{ab}$ is the Schouten tensor and $K_{abc}{}^d := \Theta^{-1}C_{abc}{}^d$ is the gravitational field tensor. In this form of the conformal field equations the conformal factor $\Theta$ is taken to be an unknown of the system, and hence the location of the conformal boundary is not known before a numerical evolution is performed, it must be determined during the evolution.

The GCFE arise by utilizing the full conformal freedom available. This is accomplished by freeing the connection from being compatible with a metric and instead demanding only that it be compatible with the conformal class of the physical metric i.e., by using a \emph{Weyl connection} $\hnabla$. This allows for the use of conformal geodesics, curves which are defined with respect to a Weyl connection in a similar way to how geodesics are defined in terms of a metric connection. The conformal Gauß gauge can be introduced in analogy to the standard Gauß gauge but being adapted to time-like conformal geodesics instead of time-like geodesics. This gives rise to an additional term in the geodesic deviation equation, which helps avoiding the development of caustics, and one can in fact obtain a semi-global covering of the Schwarzschild space-time in this gauge \cite{Friedrich:2003}.

A startling consequence of exploiting the full conformal freedom through introducing a Weyl connection and the conformal Gauß gauge is that the conformal factor can now be explicitly determined from initial data alone. This means that the location of $\scri$ is known before the evolution proceeds. In addition, the evolution equations simplify significantly which has important consequences for the design of boundary conditions. We will now briefly describe the system of equations and the variables we use for the problem at hand.

In complete analogy to other approaches in numerical relativity we perform a $3+1$ decomposition of the space-time and the variables. The difference in our approach compared to others is due to the use of spinors to decompose the variables into irreducible parts. This is motivated mainly by the fact that the Bianchi equation for the rescaled Weyl spinor takes a very efficient form when written in the spinor formalism compared to a tensorial formulation. We use the space-spinor formalism which implements the $3+1$ decomposition in a straightforward way. Most tensorial quantities are written in a spinorial representation with respect to a spin-frame $(o^A,\iota^A)$ and its complex conjugate. The resulting $SL(2,\CC)$ spinors are then transformed to $SU(2)$ spinors i.e., primed indices are transformed to unprimed ones using the Hermitian form $t_{AA'}$ (normalised by $t_{AA'}t^{AA'}=2$) on spin space defined by the time-like tangent vector to the congruence of time-like conformal geodesics used for the Gauß gauge. Finally, the resulting unprimed spinors are then decomposed into irreducible parts.

In a similar way, we split the covariant derivative $\nabla_a$ at every point of the space-time manifold into a part which is parallel to $t^a=t^{AA'}$, and a part orthogonal to it. We define $\del$ as the covariant derivative operator in the tangential direction which annihilates $t^a$ and $\del_{AB} = \del_{(AB)}$ is a covariant derivative in the orthogonal direction, also annihilating $t^a$, so that 
\begin{equation}
	t_B{}^{A'}\nabla_{AA'} = \frac12\eps_{AB}\del + \del_{AB}.
\end{equation}
We define frame components $c^\mu_{AB}$ as the action of $\del_{AB}$ on the coordinates
\begin{equation}\label{eq:delABexpansion}
	c^\mu{}_{AB} := \del_{AB}x^\mu,
\end{equation}
and the $\gamma_{ABC}$ as the action of $\del_{AB}$ on the spin-frame
\begin{equation}
	\gamma_{ABC} := \del_{AB}o_C,
\end{equation}
where the analogous quantities defined with the $\del$ are fixed as $c^\mu := \del x^\mu = \sqrt{2}\delta^\mu{}_0$ and $\gamma_A := \del o_A=0$.

Written in the form of \cite{Beyer:2017} the GCFE are given as a system of differential equations for the unknowns
\begin{equation}\label{eq:systemvars}
(c^\mu_{AB},\, \gamma_{ABC},\, K_{ABCD},\, f_{AB},\, P_{ABCD},\, \psi_{ABCD}),
\end{equation}
where $c^\mu_{AB}$ are frame components, $\gamma_{ABC}$ are connection coefficients, $K_{ABCD}$ can be thought of as the extrinsic curvature of hypersurfaces of constant time, $f_{AB}$ is defined by the action of the Weyl connection $\hnabla$ on the conformal metric $g$ as $\hnabla_c g_{ab} = -2f_cg_{ab}$, $P_{ABCD}$ is the Schouten spinor with respect to the Weyl connection and $\psi_{ABCD} := \Theta^{-1}\Psi_{ABCD}$ is the rescaled Weyl spinor, which describes the gravitational field.

These are evolved with the symmetric hyperbolic evolution system 
\begin{subequations}\label{eveqs}
\begin{align}
  \del c^0_{AB} &= -\sqrt{2}f_{AB} - K_{AB}{}^{CD}c^0_{CD}, \label{ev:6} \\
  \del c^i_{AB} &= -K_{AB}{}^{CD}c^i_{CD}, \qquad i=1,2,3. \label{ev:7} \\
  \del K_{ABCD} &= -K_{AB}{}^{EF}K_{EFCD}  - 2P_{AB(CD)} + \Theta\psi_{ABCD} + \Theta\hat\psi_{ABCD}, \label{ev:3} \\
  \del\gamma_{ABC} &= -K_{AB}{}^{EF}\gamma_{EFC} - o_{(A}K_{B)CDE}f^{DE} + \frac12o_CK_{ABEF}f^{EF} +
  K_{ABE(C}f_{D)}{}^Eo^D \nonumber \\
  &\hskip6em + \frac12P_{ABE}{}^Eo_C + \eps_{C(A}P_{B)DE}{}^Eo^D - \frac12\Theta\psi_{ABCD}o^D + \frac12\Theta\hat\psi_{ABCD}o^D , \label{ev:5} \\
  \del f_{AB} &= -K_{ABEF}f^{EF} + P_{ABC}{}^C, \label{ev:4} \\
  \del P_{ABCD} &= -K_{ABEF}P^{EF}{}_{CD} + \psi_{ABCE}h^E{}_D - \hat\psi_{ABDE}h_C{}^E, \label{ev:1} \\
  \del\psi_{ABCD} &= 2\del_{(A}{}^E\psi_{BCD)E} - 2K_{(A}{}^E\psi_{BCD)E} + 3K_{(A}{}^E{}_B{}^F\psi_{CD)EF} - K_{E(A}{}^{EF}\psi_{BCD)F}. \label{ev:2} 
\end{align}
\end{subequations}
Here, $\hat\psi_{ABCD}$ denotes the Hermitian conjugate of $\psi_{ABCD}$ defined in the space-spinor formalism by complex conjugation followed by transvection with appropriately many $t^{A'}{}_A$. It can be seen from this system that there are no equations for the conformal factor $\Theta$ and the 1-form $h_a$. In fact, the conformal factor and the components of $h_a$ can be obtained explicitly from the Gauß gauge \begin{gather}
	\Theta(t ) = \ul{\Theta}  + \ul{Z} t  + \frac14 \ul{H}\, t^2, \\
	h_{0}(t ) = \frac{1}2 \ul{H}t  + \ul{h}_{0},\qquad
  	h_{k}(t ) = \ul{h}_{k},\qquad k=1,2,3,
\end{gather}
where the underlined quantities are computed solely with initial data at $t =0$. The corresponding constraint equations are given in the appendix.

As indicated above this system of evolution equations shows a remarkable structure: all equations except for~\eqref{ev:2} are simply transport equations along the time-like conformal geodesics. Only the equation for the gravitational field features a derivative in spatial directions. In fact, it is easy to show that this subsystem for the rescaled Weyl spinor is symmetric hyperbolic. This fact is mirrored in the constraint propagation system (see \cite{Beyer:2017} for these equations) in which the constraints for $\psi_{ABCD}$ are propagated with a symmetric hyperbolic system of equations as well, while all the other constraint quantities are simply advected  along the time-like conformal geodesics.

Having defined the system, we need to take care of imposing boundary conditions for the components of $\psi_{ABCD}$. This has to be dealt with very carefully, and we follow the maximally dissipative approach as in \cite{Friedrich:1999}, which was successfully numerically implemented also in \cite{Frauendiener:2014b}. This approach looks at how an ``energy'' changes over time, and the boundary conditions are chosen so that no energy propagates in from the boundaries. In short, using that the evolution system for $\psi_{ABCD}$ is symmetric hyperbolic we can diagonalize it at every boundary point and find the five distinct characteristic variables $\{\tpsi_0,\tpsi_1,\tpsi_2,\tpsi_3,\tpsi_4\}$ with speeds $\{-\alpha,-\beta,0,\beta,\alpha\}$, $\alpha>\beta>0$ perpendicular to the boundary. These speeds correspond to $\tpsi_0$ and $\tpsi_4$ travelling along light-cones and $\tpsi_1$ and $\tpsi_3$ travelling in a time-like fashion. The diagonalization procedure yields a transformation between the $\psi_i$ and the diagonalized $\tpsi_i$. One can prescribe free data for those components of $\tpsi_i$ that travel inwards. These resulting fields are then translated back into boundary conditions for the components of $\psi_i$.

The structure of the Weyl system~\eqref{ev:2} shows that at a given boundary there will be two ingoing modes, two outgoing modes and one mode travelling tangential to the time-like boundary. The boundary conditions must be such that they do not inject any constraint violating modes into the domain because these would cause the violation of the constraints across the entire domain and hence render the computation useless.

Thus we must also perform a characteristic analysis of the constraint propagation system. The only constraint equation of interest is
\begin{equation}
	G_{AB} := \del^{CD}\psi_{ABCD} + K^{CE}{}_E{}^D\psi_{ABCD} + K^{CDE}{}_{(A}\psi_{B)CDE} = 0,
\end{equation}
whose propagation equation has the principal part
\begin{equation}
	\del G_{AB}=\del^C{}_{(A}G_{B)C}.
\end{equation}
The characteristics of the components of this constraint propagation equation for the components $\{G_0,G_1,G_2\}$ are $\{-\beta,0,\beta\}$. It is of no surprise that there is a correspondence between the characteristics of the evolution and subsidiary systems, see for example~\cite{Alcubierre:2008}. These equations have one ingoing mode which is the one that needs to be suppressed at the boundary. The vanishing of this mode leads to a condition on the ingoing modes for the $\psi_i$ which eliminates one degree of freedom in the choice of boundary data. Thus, at every boundary there is exactly one degree of freedom, i.e., one complex valued function, that can be specified freely on the boundary. It corresponds exactly to the ingoing spin-2 component of the Weyl spinor with respect to the boundary. We do not elucidate this procedure further here and again refer to \cite{Beyer:2017} for a full account.

\section{The Bondi-Sachs components}
\label{sec:Bondi4momentum}

In this section we summarize the results obtained in~\cite{Frauendiener:2021a} and put forward the analytical expressions required to compute the Bondi-Sachs energy on arbitrary cuts of $\scri^+$.

First, we need to transform to a \emph{Bondi frame}, say $\{L^a,N^a,M^a,\overline{M}^a\}$, which is a null tetrad such that $N^a$ points along the null generators\footnote{Note, that here we divert in notation with \cite{Penrose:1986} where $N^a=-\nabla^a\Theta$.}, and that $M^a$ is tangent to a given cut~$C$. These conditions are expressed as equations on $\scri^+$
\begin{equation}\label{eq:BFConditions}
	\nabla_a\Theta = -A N_a,\qquad
	M^a\nabla_a t  = 0,
\end{equation}
where $A$ is a positive non-zero scalar on $\scri^+$. The inclusion of this factor is necessary for maintaining the GHP formalism. However, even if one would fix the choice of a frame in some way then $A$ would show up as a normalisation factor. This frame can be thought of as a ``physical detector'' on $\scri^+$, which can ``detect'' the radiation coming from the physical space-time.

Next we consider the Bondi-Sachs momentum 4-covector components, henceforth called Bondi components, where each component is defined as an integral over the cut~$C$. The work in~\cite{Frauendiener:2021a} results in a generalised expression for these components and represents them in a conformally invariant GHP (cGHP) formalism. As there are many quantities in these equations that need an explanation, we first present them all and then unpack their meanings. The Bondi components are obtained from integration over $C$ through
\begin{equation}\label{eq:BondiIntegral}
	 m_C[U] = \frac1{4\pi}\int_{C}U\Big{(}\sigma N + \overline{\eth}_c^2 \sigma - A\psi_2\Big{)}A^{-1}\mathrm{d}^2\Sigma,
\end{equation}
where
\begin{equation}\label{eq:DefnOfN}
	N := \Phi_{20} - \rho'\bar{\sigma} - \eth'\bar{\tau} + \bar{\tau}^2,
\end{equation}
is closely related to Bondi's \emph{news function} and $U$ is an asymptotic translation obtained from the equations
\begin{equation}\label{eq:Ueqs}
	\eth_c^2U = \cR U, \qquad 
	\overline{\eth}_c\cR + \eth_c\cQ = 0, \qquad
	\cQ := K - \eth'\tau.
\end{equation}
To unpack all this, we note the presence of the spin-coefficients $\rho', \sigma, \tau$ from the Newman-Penrose formalism, the $\eth$ and $\eth'$ operators from the associated GHP formalism and their conformally invariant counterparts $\eth_c,\overline{\eth}_c$ from the cGHP formalism. The quantity $\Phi_{20}$ is a component of the spinor $\Phi_{ABA'B'}$ which essentially represent the trace-free Ricci tensor and $\psi_2$ is the spin-zero component of the gravitational spinor. Finally, $K$ is related to the Gauß curvature $k$ of the cut which is given on $C$ by $k := 2K = 2(\Phi_{11} + \Lambda - \rho\rho')$, which is a real quantity, where $\Lambda$ is the scalar curvature and $\Phi_{11}$ is another component of $\Phi_{ABA'B'}$. We note that the three terms in the parentheses in Eq.~\eref{eq:BondiIntegral} are together referred to as the \emph{mass aspect}.

Eq.~\eref{eq:BondiIntegral} is manifestly conformally invariant, assumes no conditions on the conformal factor $\Theta$, which is used to compactify the space-time,  or spin-coefficients and reduces to the original definition when the specialization in \cite{Penrose:1986} is taken. $U$ represents an infinitesimal translation. The translations form a subgroup of the super-translation group, which itself is a subgroup of the BMS group. The quantity $\cQ$ arises naturally from action of the commutator $[\eth_c,\overline{\eth}_c]$ on properly weighted quantities and the quantity $\cR$ is uniquely determined by $\cQ$ and is referred to as the \emph{co-curvature}.

If the cut was a unit sphere and if we were in standard polar coordinates, then the equation for $U$ would reduce to $\eth^2U=0$ and it would have four independent solutions, namely the first four spherical harmonics $Y_{lm}$ ($l=0,\; m=0$ and $l=1,\; m=-1,0,1$.) These four choices for $U$ correspond to asymptotic time and space translations, and when normalized with a certain Lorentzian metric lead to the energy and momenta, respectively, when integrated against the mass aspect as given by Eq.~\eref{eq:BondiIntegral} in a Bondi frame.

It is useful to present the Bondi-Sachs mass-loss
\begin{equation}\label{eq:BSML}
	m_2[U] - m_1[U] = -\frac{1}{4\pi}\int_{\scri_1^2}\mathcal{N}\overline{\mathcal{N}}\text{d}^3\text{V},
\end{equation}
where $\mathcal{N}:= N + \bar{\mathcal{R}}$ and $\scri_1^2$ is the part of null infinity between two cuts $C_1$ and $C_2$ on which the corresponding component is evaluated. When $U$ represents an appropriately normalized asymptotic time translation, Eq.~\eref{eq:BSML} relates the difference in Bondi energy at two different cuts of $\mathscr{I}^+$ to the ``energy flux'' between them due to the outgoing gravitational radiation. This will be numerically checked in Sec.~\ref{sec:BSML}.

In this paper we consider only the Bondi energy, where the associated $U$ is equal to the conformal factor $\Omega$ that scales the metric of a given cut $C$ of $\scri^+$ to the unit 2-sphere metric~\cite{Frauendiener:2021a}.

\section{The setup}
\label{sec:IDandBD}

In the previous section we outlined how to solve an IBVP for the GCFE, taking proper care of the boundary conditions. We now discuss our choice of initial data. We choose to evolve Schwarzschild initial data from a space-like initial hypersurface as laid out in \cite{Friedrich:2003} and following the previously implemented numerical evolution in \cite{Beyer:2017}. This choice of initial data, evolved with the GCFE, was analytically shown~\cite{Friedrich:2003} to remain smooth and free of degeneracies up to and including null infinity. Even with gravitational waves impinging on this space-time region from the time-like outer boundary, we have demonstrated in~\cite{Beyer:2017} that the evolution of these initial data remains valid up to $i^+$, where the code breaks down due to the physical singularity there.

Even though the code is written to handle a general 3D problem for efficiency we now assume that the space-time is axisymmetric to reduce the problem to $2+1$ dimensions.
This has the consequence that the spectral coefficients of a function in the spin-weighted spherical harmonic basis ${}_sY_{lm}$ vanish when $m=0$.

We assume that the initial hypersurface can be foliated into concentric spheres and we adapt the coordinate system to this choice by labeling each sphere with a radial coordinate $r$ and introducing standard polar coordinates $(\theta,\phi)$ on each sphere. We define complex derivative operators on each sphere by
\begin{equation}
	\tdelta = \frac{1}{\sqrt{2}}\Big{(}\del_\theta - \frac{\ii}{\sin\theta}\del_\phi\Big{)},\qquad
	\tdelta' = \frac{1}{\sqrt{2}}\Big{(}\del_\theta + \frac{\ii}{\sin\theta}\del_\phi\Big{)}.
\end{equation}
By deeming these to be complex null vectors we have implicitly introduced a fiducial metric on each sphere which makes it into a unit-sphere. This has nothing to do with the actual metric on each sphere induced from the space-time metric and exists for the sole purpose of having available a framework for using a numerical $\eth$-formalism based on the unit-sphere.

This structure on the initial hypersurface is carried along by the evolution along the conformal geodesics which are parameterised by the coordinate $t$. Thus, we use as a basis in each tangent space  the vectors $\{\del_t , \del_r, \tdelta, \tdelta'\}$ with dual basis $\{\text{d}t ,\text{d}r,\bftm, \bftmb\}$, where $\{\tdelta,\,\tdelta'\}$ are tangent to the spheres of constant $(t,r)$. We assume the range of radial values lie in an interval so that the left and right time-like boundaries at a fixed $t$ will be spheres of constant $r$, denoted by $r_l$ and $r_r$ respectively with $r_l < r_r$. 

\subsection{Initial data}
The Schwarzschild metric in isotropic coordinates is
\begin{equation}\label{eq:SchwarzschildLineElement}
	\tg = \Big{(}\frac{1-\frac{m}{2r}}{1+\frac{m}{2r}}\Big{)}^2\text{d}\tilde{t}^2
	-(1 + \frac{m}{2r})^4\Big{[}\text{d}r^2 + r^2\Big{(}\text{d}\theta^2
	+ \sin^2\theta\,\text{d}\phi^2\Big{)}\Big{]}.
\end{equation}
The initial data to fix the conformal Gau\ss\;gauge on the surface $t=0$ are chosen following \cite{Friedrich:2003}. An underline is used to represent functions evaluated there. We first fix the conformal factor on the surface as 
\begin{equation}
	\ul{\Theta} = \frac{r^2}{(r + \frac{m}{2})^4}.
\end{equation}
This then defines a conformal metric $\ul{g} = \ul{\Theta^2}\ul{\tg}$, a frame $\{\ul{e_0},\ul{e_1},\ul{m},\ul{\bar{m}}\}$
\begin{gather}
	\ul{e_0} = \frac{\Big{(}r + \frac{m}{2}\Big{)}^5}{r^2\Big{(}r - \frac{m}{2}\Big{)}}\partial_{\tilde{t}}, \qquad
	\ul{e_1} = \Big{(}r + \frac{m}{2}\Big{)}^2\partial_r, \qquad
	\ul{m} = \frac{\Big{(}r + \frac{m}{2}\Big{)}^2}{r}\tdelta.
\end{gather}
which satisfies $-\ul{g}(\ul{e_0},\ul{e_0}) = \ul{g}(\ul{e_1},\ul{e_1}) = \ul{g}(\ul{m},\ul{\bar{m}}) = -1$ and we choose $\ul{h}_a = \ul{\nabla_a}\ul{\Theta}$ to fix the remaining freedom in the gauge. This yields the full expressions for $\Theta$ and $h_a$ as
\begin{gather}
	\Theta = \frac{r^2}{\Big{(}r+\frac{m}{2}\Big{)}^4} 
	- t^2\Big{(}\frac{r - \frac{m}{2}}{r + \frac{m}{2}}\Big{)}, \label{eq:CFforSchwarzschild}\\[4pt]
	h_0=h_2=0,\quad h_1 = -\sqrt{2}r\frac{r - \frac{m}{2}}{\Big{(}r + \frac{m}{2}\Big{)}^3},\quad
	h = -2\sqrt{2}t \Big{(}\frac{r - \frac{m}{2}}{r + \frac{m}{2}}\Big{)}^2,\label{eq:hforSchwarzschild}
\end{gather}
where we have decomposed $h_a$ as $(1/2)h\,\eps_{AB} + h_{AB}$, with $h_{AB}=h_{(AB)}$. We find initial data for the system variables \eqref{eq:systemvars} by using together the evaluation of the line element \eqref{eq:SchwarzschildLineElement} at $t=0$, \eqref{eq:CFforSchwarzschild}, \eqref{eq:hforSchwarzschild} and the condition that $K_{ABCD}=0$ at $t=0$. The result is the initial data set of non-vanishing system variables as
\begin{gather}
	\ul{c^2{}_0} = -\ul{c^3{}_2} = -\frac1r\Big{(}r + \frac{m}{2}\Big{)}^2,\qquad 
	\ul{c^1{}_1} = \frac1{\sqrt2}\Big{(}r + \frac{m}{2}\Big{)}^2,\qquad
	\ul{\gamma_{20}} = \ul{\hat{\gamma}_{01}} = \frac1{\sqrt{2}r}\frac{r + \frac{m}{2}}{r - \frac{m}{2}},\nonumber \\
	\ul{P_{101}} = \ul{P_{110}} = \frac{m}{r}\Big{(}r + \frac{m}{2}\Big{)}^2,\qquad
	\ul{\psi_2} = -\frac{m}{r^3}\Big{(}r + \frac{m}{2}\Big{)}^6,\label{eq:SchwarzschildID}
\end{gather}
where $\hat{\gamma}_{ABC}$ is the complex conjugate of $\gamma_{ABC}$ as defined in \cite{Beyer:2017}.

\subsection{Boundary data}
At this point, \eqref{eq:CFforSchwarzschild}, \eqref{eq:hforSchwarzschild} and \eqref{eq:SchwarzschildID} represent exact Schwarzschild initial data together with a specific choice of the conformal Gauß gauge. Now we need to choose our boundary conditions on $r_l$ and $r_r$. As discussed in Section \ref{sec:OverviewOfIBVP}, the characteristics for $\{\tpsi_0,\tpsi_1,\tpsi_2,\tpsi_3,\tpsi_4\}$ have signs $\{-,\,-,\,0,\,+,\,+\}$. Thus we must provide boundary data for $\tpsi_3$ and $\tpsi_4$ on the $r_l$ boundary and provide boundary data for $\tpsi_0$ and $\tpsi_1$ on the $r_r$ boundary. These translate into the boundary conditions for the corresponding untilded system variables that are used in the evolution. As in Section \ref{sec:OverviewOfIBVP} we fix $\tpsi_1$ and $\tpsi_3$ uniquely by imposing that the ingoing mode of the constraint propagating system vanishes, thus being left with prescribing the physically relevant $\tpsi_0$ and $\tpsi_4$. We do this by choosing the free datum $q_0$ for $\tpsi_0$ and $q_4$ for $\tpsi_4$, which represent the free wave profiles, as
\begin{align}
	q_0(t ,r_r,\theta) &=
	\begin{cases} 
	    4a\mathrm{i}\sqrt{\frac{2\pi}{15}}\;{}_2Y_{20}(\theta)\sin^8(4{\pi t })& t \leq\frac14 \nonumber \\
	    0 & t >\frac14
	\end{cases}, \\[6pt]
	q_4(t ,r_l,\theta) &= 0,\label{eq:BCs}
\end{align}
where the constant $a$ is the amplitude of the ingoing gravitational wave. We note that the choice of an imaginary wave profile has the welcome side effect of allowing the system variables to become complex which gives rise to more checks of correctness.

\subsection{Numerical setup}
We implement the above IBVP in the Python package COFFEE \cite{Doulis:2019} which contains all the necessary infrastructure as previously described in~\cite{Beyer:2017}. As a brief overview of the numerical scheme, we use: the method of lines, an explicit RK4 method for time integration, a fourth order finite differencing scheme (third order close to the boundaries) that satisfies the summation-by-parts property for approximating derivatives in the $r$-direction \cite{Strand:1994}, a fast spin-$s$ spherical harmonic transform for approximation of the $\eth$ and $\eth'$ derivatives \cite{Huffenberger:2010} and the simultaneous approximation term method for stable imposition of boundary conditions \cite{Carpenter:1994}.

Fixing the Schwarzschild mass as $m=1/2$, we discretized the $r$ and $\theta$ directions into equidistant points in the two-dimensional interval $[m/2, 5m/2]\times[0,\pi]$. Convergence tests were carried out in \cite{Beyer:2017} and showed that the constraints propagate and converge at the correct order; we refer the reader there for more details. Thus, we can be confident that the code produces a converging approximation to the solution of the IBVP.

\subsection{Regridding}

Ideally we want to evolve our fields as far along $\scri^+$ toward time-like infinity $i^+$ as possible. However, due to the nature of the time-like conformal geodesics, on which our evolution scheme is based, as we get closer to $i^+$ an increasing portion of the computational domain lies either inside the black hole  horizon or outside $\scri^+$. The region inside the horizon will inevitably hit the singularity, causing system variables to diverge and stop our simulation. In addition, the evolution does not behave well in the unphysical region outside $\scri^+$  since the $t =\const$ surfaces tend to become null, which creates large gradients and ultimately causes the simulation to crash.

To remedy this, we simply cut these parts out of the computational domain once certain conditions are met, such as fields getting too large when approaching the singularity or $\scri^+$ has gotten a certain number of radial grid points away from the outer boundary. Neither of these regions are of interest to us, and both regions cannot physically influence the domain between them including $\scri^+$. 

Suppose the regridding procedure has been activated on the left boundary, then the simulation proceeds as follows: The current time $t_c$ and all system variables are written out to NumPy files. The spatial computational domain is then cut at a fixed $r_c\in(r_l,\,r_r)$, so that the new radial interval becomes $[r_c,\,r_r]$. A new computational grid is created by discretizing the new interval into the same number of points as the old interval. The system variables are then interpolated onto the new interval's grid points using a B-spline of order 5 with the Python package SciPy. These are then fed back into COFFEE as initial data at time $t_c$ and the simulation continues.

After the first regrid, one has to be careful as to what boundary conditions are imposed. Take the outer boundary as the example. Suppose we have evolved our system to a time after the wave has been introduced and the boundary is outside $\scri^+$. Then the boundary conditions are $\psi_0=0$ from Eq.~\eref{eq:BCs} and the other for $\psi_1$ is fixed by the procedure outlined in Sec.~\ref{sec:OverviewOfIBVP}. Now suppose the first regrid has occurred. What should the boundary condition be on the \emph{new} outer boundary, which used to be an inner point? It is unlikely that the $\psi_0=0$ boundary condition is compatible with whatever the value of $\psi_0$ already is there and this could potentially introduce an instability, or at best, ``kinks'', into the system variables due to the violation of corner conditions.

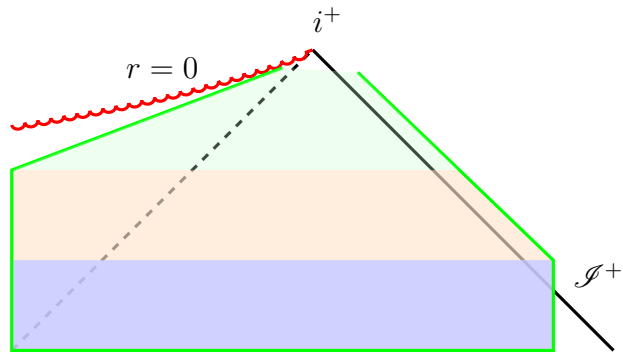
\begin{figure}[h]
    \centering
    \begin{tikzpicture}[very thick, decoration = {bent,amplitude=-5}, scale=2]
        \draw[dashed] (0,0) -- (2,2);
        \draw (2,2) -- (4,0);
        \draw[red,decoration={bumps,amplitude=-2}, decorate] (0,1.5) to[out=10, in=200] (2,2);
        \node at (2.1, 2.2) {$i^+$};
        \node at (3.9, 0.5) {$\mathscr{I}^+$};
        \node at (1, 1.88) {$r=0$};
        \fill[fill=blue!20!white, fill opacity=0.9]
        (0, 0.6) -- (0,0) -- (3.6,0) -- (3.6,0.6);
        \fill[fill=orange!20!white, fill opacity=0.65]
        (0,1.2) -- (0, 0.6) -- (3.6,0.6) -- (2.97, 1.2);
        \draw[draw  = green, very thick]
        (1.8, 1.88) -- (0,1.2) -- (0,0) -- (3.6,0) -- (3.6,0.6) -- (2.3, 1.85);
        \fill[fill=green!20!white, fill opacity=0.3]
        (1.8, 1.88) -- (0,1.2) -- (2.97, 1.2) -- (2.3, 1.85);
      \end{tikzpicture}
      \caption{A diagram showing how the computational domain changes with the regridding procedure. The first region (blue) contains no regridding. The middle region (orange) is when regridding occurs outside $\mathscr{I}^+$. The top region (green) is where regridding also occurs inside the black hole.}
      \label{fig:RegriddingDiagram}
\end{figure}

Both of our boundaries lie in regions where information cannot propagate into the physical space-time outside the black hole. Thus one only needs a numerically stable way of handling the boundaries. A resolution to this problem, for both inner and outer boundaries, is to stop imposing boundary conditions at all. This is numerically reasonable as the evolution procedure as the method of lines will always fill these boundary points. This procedure is found to be numerically stable. It does affect the constraints numerically on $\mathscr{I}^+$, but in a small enough way to keep the subsequent results valid. Regridding occurs first on the right boundary as this always passes beyond $\mathscr{I}^+$ before the left boundary propagates close to the singularity. Fig.~\ref{fig:RegriddingDiagram} shows how the computational domain will change with regridding and Fig.~\ref{fig:constraint-diff-amps} from Sec.~\ref{sec:Results} shows the effect along $\mathscr{I}^+$ and in the physical space-time arising from wave packets of different amplitudes.

\section{Calculating the Bondi components}\label{sec:CalculatingtheBondiComponents}

\subsection{Data on \texorpdfstring{$\scri^+$}{null infinity}}

The first step in preparing ourselves to compute the Bondi components is to locate $\scri$ in the computational domain. In \cite{Beyer:2017} it was shown, using the choice $m=1/2$, that the numerical evolution will eventually reach null infinity and even go through it into the unphysical region. Once this has happened then on a $t =\const$ time-slice that intersects $\scri^+$, the cut is found at a fixed and analytically known $r$ value due to the conformal factor being independent of the angular coordinates. The values of the system variables on this cut are then computed via interpolation. Doing this on each $t =\const$ surface that contains $\scri^+$ gives us a sequence of cuts and values of the system variables on them. Below, we will need first and second derivatives of the system variables evaluated on $\scri^+$. Time derivatives can be computed through use of the evolution equations, radial derivatives can be computed using information from within the time-slice and then interpolating to $\scri^+$ and the angular derivatives can be computed directly on $\scri^+$. 

\subsection{Transforming to a Bondi frame}\label{sec:trafo-to-Bondi-frame}

The next step is to set up a null tetrad that is adapted to $\scri^+$ and the cuts. We will denote the adapted frame by $\{L^a,N^a,M^a,\overline{M}^a\}$ with corresponding spin-frame $\{O^A,I^A\}$. This can be accomplished through null rotations of the original spin-frame. We first rotate $\iota^A$ around $o^A$ to align it with the null generators of $\scri$, and then rotate $o^A$ around the new $\iota^A$ to make $M^a$ tangent to the cuts. This gives
\begin{equation}\label{eq:spinframe}
	O^A = o^A + Y(\iota^A + X o^A),\qquad I^A = \iota^A + Xo^A.
\end{equation} 
The null rotation functions $X,Y$ are fixed by the requirements that $\nabla_a\Theta = -AN_a$ and $M^a\nabla_a t = 0$ which yield expressions involving frame coefficients and derivatives of the conformal factor.


It is noted that for our choice $\gamma_A=0$ the spin-frame $\{o^A,\,\iota^A\}$ automatically satisfies that $m^a = o^A\iota^{A'}$ is tangent to each cut of $\scri^+$, so that $X=0$.

The next step is to calculate the necessary spin-coefficients and the Ricci spinor component in this frame as well as transform $\psi_2$. We redefine quantities such as $\psi_2$, $\rho'$  to mean quantities with respect to the adapted frame to avoid cluttered notation. Then
\begin{align}\label{eq:BFexprs}
	\psi_2 &:= \psi_{ABCD}O^AO^BI^CI^D, \nonumber \\
	\sigma &:= O^AO^BI^{B'}\nabla_{BB'}O_A, \nonumber \\
	\rho'  &:= -I^AO^BI^{B'}\nabla_{BB'}I_A, \nonumber \\
	\Phi_{20} &:= \Phi_{ABA'B'}I^AI^B\bar{O}^{A'}\bar{O}^{B'}.
\end{align}
These are calculated by using \eref{eq:spinframe} to replace the Bondi spin-frame spinors $O^A,I^A$ with $X,Y,o^A,\iota^A$ so the expressions are given in terms of system variables, null rotation functions, and their derivatives. Using this approach, any spin-coefficient or curvature component can be obtained with respect to the new null tetrad.

\subsection{Transformation of the \texorpdfstring{$\eth$}{eth}-operators}
\label{sec:transform-eth}

The last and most awkward task is to express the $\eth$-derivative as used in Sec.~\ref{sec:Bondi4momentum} in terms of the derivatives with respect to the coordinates $(t,r)$ and angular derivatives encoded in the numerical $\tilde\eth$-operators. The action of any $\eth$-operator on a function $f$ with weights $(p,q)$ is given by
\begin{equation}
	\eth f = (\delta - p\beta + q\bar{\beta}')f,\qquad
	\eth f = (\delta' + p\beta' - q\bar{\beta})f,
\end{equation}
where the $\delta$-derivatives and the spin-coefficients are computed with respect to some complex null vector $m^a$. For the Bondi $\eth$-operators these are the vectors $M^a$ and $\overline{M}^a$.

In order to calculate the action of these operators we reexpress the quantities appearing on the right-hand side with those that we have available, namely the GCFE system variables, $t $ and $r$ coordinate derivatives and the numerical $\teth$-derivatives. The spin-coefficients can be reexpressed in the same way as done in \eref{eq:BFexprs}, and the operators $\delta$, $\delta'$ are easily reexpressed via
\begin{align}
	\delta = M^a\nabla_a = F_t \del_t  + F_{r}\del_r + F_{\delta}\tdelta + F_{\delta'}\tdelta', \\
	\delta' = \overline{M}^a\nabla_a = \bar{F}_t \del_t  + \bar{F}_{r}\del_r + \bar{F}_{\delta'}\tdelta + \bar{F}_{\delta}\tdelta',
\end{align}
where the $F_i$ are known functions of the null rotation functions $X$, $Y$ and the frame coefficients $c^\mu{}_i$. 
In the same way one can express the Bondi frame $\eth$ operators in terms of the numerical $\teth$ operators by using the appropriate connection coefficients $\gamma_{ABC}$. This process results in properly weighted equations so that all equations can be expressed consistently with this $\eth$-formalism.

\subsection{Coordinate expression of the area element}
\label{sec:coord-trafo-2sphere}

The area element $\text{d}^2\Sigma$ used in~\eqref{eq:BondiIntegral} is the one given with respect to the induced metric on the cut $C$. Thus, we can compute it as the pull-back of the 2-form $2\mathrm{i}\overline{M}_{[a}M_{b]}$ to the cut and it must be proportional to the coordinate 2-form $\mathrm{d}^2S := \sin\theta\, \mathrm{d}\theta \wedge \mathrm{d}\phi$. Using the expansion of $M_a$ in terms of the coordinate differentials and the null rotation functions we obtain
\begin{equation}
  \text{d}^2\Sigma = J\,\text{d}^2S,
\end{equation}
where $J$ is a scalar function on the cut, a known expression involving the frame components and the null rotation functions.

\subsection{Conformal rescaling to the unit 2-sphere}
\label{sec:conformal-trafo-2sphere}

As mentioned in Sec.~\ref{sec:Bondi4momentum}, to compute the Bondi energy on a cut one needs an appropriate time translation $U$. This can be chosen as the conformal factor which scales the unit-sphere to the induced metric on the cut. This is not the only possibility, other choices correspond to frames which are relatively boosted and therefore give a different energy component. In principle, one should be concerned with the invariant mass of the system, but in the present case it is easy to see that due to the axisymmetry, any momentum contribution must lie along the symmetry axis, and due to the initial and boundary data being invariant under reflection at the equator also this component will vanish. Thus, the energy is equal to the mass in the present context.

In order to compute the conformal factor $\Omega$ one needs to consider the transformation of the Gauß curvature $k$ of a surface under the rescaling of its metric~$q_{ab}$. Let $q_{ab} = \Omega^2\hat{q}_{ab}$ then the corresponding curvatures are related by 
\begin{equation}\label{eq:Omega-to-unit-sphere}
	\hat{k} = k\Omega^2 + \Omega\nabla^a\nabla_a\Omega - \nabla_a\Omega\nabla^a\Omega,
\end{equation}
where $\nabla_a$ denotes the covariant derivative with respect to the metric $q_{ab}$. In our context, this metric is the induced metric $2M_{(a}\overline{M}_{b)}$ on the cut $C$ and $\nabla_a$ can be expressed in terms of the $\eth$-operators defined with respect to the complex null vector $M^a$ which is tangent to the cut. The unit-sphere has Gauß curvature $\hat{k}=1$ so that the equation reads
\[
  k\Omega^2 - 2\Omega\eth\eth'\Omega + 2 \eth\Omega\,\eth'\Omega = 1.
\]
The Gauß curvature $k$ of the induced metric on the cut can be obtained from the 4-dimensional curvature since $k = K + \bar{K}$ where $K$ is the ``complex curvature''
\[
  K = \sigma\sigma'  - \Psi_2 + \Phi_{11} + \Lambda - \rho\rho',
\]
which simplifies on $\scri$, where both $\sigma'$ and $\Psi_2$ vanish, so that
\[
  k = 2(\Phi_{11} + \Lambda - \rho\rho').
\]
Expressing the $\eth$-derivatives as before in terms of the numerical derivatives the equation is then represented as the vanishing of
\begin{equation}
	Z_0 \teth^2\Omega + Z_1 \teth\teth'\Omega + Z_2\teth'\teth\Omega + Z_3\teth'^2\Omega + Z_4\teth\Omega + Z_5\teth'\Omega + \tilde{k}\Omega^2 - 1,
\end{equation}
where the functions $Z_i$ are known numerically on the cut and recalling that $\teth,\teth'$ are the numerical $\eth$-operators on the sphere. This equation can be solved by transforming the expression to the spin-weighted spherical harmonic basis ${}_sY_{lm}$\footnote{This describes the procedure in general. In the present case, all terms with $m\ne0$ vanish.}, using the Clebsch-Gordon coefficients to calculate products of the basis functions directly in spectral space, using the expansion $\Omega = \sum_{l,m}\Omega_{lm} Y_{lm}$ to get a non-linear algebraic system of equations for the expansion coefficients $\Omega_{lm}$, which is solved numerically. Once the spectral coefficients for $\Omega$ are known $\Omega$ can be reconstructed as a function on the cut. 
 
\subsection{Calculation of the Bondi energy}
\label{sec:calcBE}

At this stage all the quantities --- the mass aspect, the time translation, the area element --- necessary  to evaluate the integral~\eqref{eq:BondiIntegral} have been determined. When everything is inserted this integral has the form
\[
  \int_{S^2}I(\theta,\phi)\, \sin\theta\, \mathrm{d}\theta\mathrm{d}\phi
\]
for some scalar function $I(\theta,\phi)$. Writing this function as an expansion in terms of spherical harmonics $Y_{lm}$ shows that it is only the ``monopole'', i.e., the coefficient $I_{00}$ of $Y_{00}$ that contributes to the integral. Thus, the value of the integral is simply
\begin{equation}
	E = m_C[\Omega] = \frac{1}{\sqrt{4\pi}}I_{00}.
\end{equation}

\section{Useful quantities}

\subsection{Location of marginally outer trapped surface}\label{sec:MOTS}

It is useful to approximate the location of the marginally outer trapped surface (MOTS), to get a rough location of the event horizon of the black hole for visualization purposes. The idea is to determine on each time-slice the locus $\rho=0$ with $\rho'\le0$, where $\rho$ resp. $\rho'$ are convergences of the outgoing resp. ingoing null congruences emanating from the spheres given by constant $r$. This is not a genuine MOTS because the convergences do not refer to that surface itself. However, it should provide a reasonable approximation which is easier to compute. We could also determine an annulus in which the MOTS should be contained by search the smallest sphere which has one point at which $\rho=0$ and the largest sphere which is entirely trapped.

The procedure to determine these spheres is straightforward. A null frame is constructed which is adapted to the spheres of constant radius $r$ on each time-slice and then the spin-coefficients $\rho$ and $\rho'$ can be found by the usual transformation rules. We construct the null frame in three steps
\begin{itemize}
	\item A null rotation of $n^a$ around $l^a$ to satisfy $\hM^a\nabla_a t =0$.
	\item A scaling of $l^a$ and $n^a$ to make $\nabla_ t $ proportional to the new $l_a+n_a$.
	\item A spatial rotation around the surface normal (which maintains the first two conditions) to make $\hM^a\nabla_ar=0$.
\end{itemize}
As before, the resulting transformation is given as a complicated expression in terms of the frame components $c_{AB}^\mu$. Once the transformation is found, the spin-coefficients can be computed and the 2-surface with $\rho=0$ and $\rho'\le0$ can be located. 

To find the true MOTS, where the above conditions for $\rho,\rho'$ are satisfied in the frame adapted to the MOTS, an iterative method must be performed as in~\cite{Schnetter:2003}. As we are only interested in an approximation for visualization purposes for now, we leave this for future work.

\subsection{Bondi time}\label{sec:BondiTime}

The retarded time $u$, also called Bondi time, is a parameter along each null generator of $\scri^+$. It is determined in terms of the conformal factor $\Theta$ which is used to compactify the space-time by the relationship
\[
  -\nabla^a\Theta \nabla_au = 1 \iff D'u = A^{-1}.
\]
It can also be characterised in terms of the second order equation along the generators
\begin{equation}
	\thorn_c'^2u = (D' - \epsilon' - \bar{\epsilon}' - 2\rho')D'u = 0.
\end{equation}
Either of these equations can be solved numerically along a null generator of $\scri^+$  with a simple Euler step using the interpolated data.

\section{Numerical results}
\label{sec:Results}

The purpose of this section is to numerically investigate the response of the Schwarzschild black hole to the gravitational wave and how the Bondi energy along $\scri^+$ is affected. We evolve as close as possible to $i^+$, at which point we expect the evolution to break down since, in the exact Schwarzschild space-time, it is a singular point. In fact, along the time-like conformal geodesic that goes through $i^+$, $\psi_2$ takes the form~\cite{Friedrich:2003}
\begin{equation}
	\psi_2 = \frac{40m^2}{(4-5m^2t^2)^2},
\end{equation}
which diverges exactly at time-like infinity located at $t =2/(\sqrt{5}m)$. 

Unless otherwise stated, we choose the initial mass of the black hole $m=0.5$ and solve the equations using a $r$-resolution of $401$ points, a $\theta$-resolution of $33$ points with an associated $l_{max}$ of $22$, a time-step of $\Delta t  = c\,\Delta r$ with CFL number $c=0.5$. The location of $i^+$ for the exact Schwarzschild space-time can be calculated \emph{a priori} using results of Friedrich \cite{Friedrich:2003}. We have $t_{i^+} = 2/(\sqrt{5}m) \approx 1.789$ and $r_{i^+} = (3+\sqrt{5})m/4 \approx 0.655$. These should be reasonable indications for the location of time-like infinity also in the perturbed cases.

\subsection{Checks of correctness}\label{sec:checks}

\begin{figure}[H]
    \centering
    \includegraphics[width=0.5\linewidth]{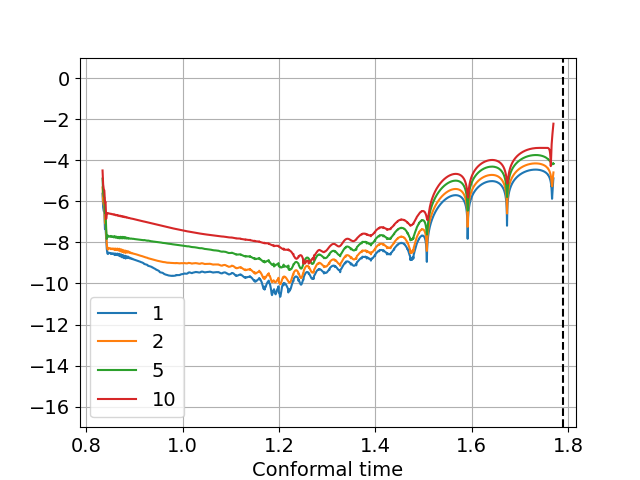}
    \caption{The effect of regridding on a constraint quantity along $\scri^+$ for a fixed angle $\theta$ and a variety of wave amplitudes $a$. The first regrid close to the right boundary occurs around $t =1.2$ and the first regrid close to the left boundary occurs around $t =1.5$. The simulations stop around $t =1.77$, the vertical dashed line represents $\mathscr{I}^+$ and the curves representing $a=1,2,5,10$ are given from bottom to top. }
    \label{fig:constraint-diff-amps}
\end{figure}

In the following list we describe details for a variety of checks done to ensure correctness of the code. In the description, an equation being ``satisfied'' is taken to mean that it converges at the correct order.
\begin{itemize}
\item All 48 components of the constraints Eq.~\eref{constreqs} converge at the correct order throughout the entire computational domain, with some of them being trivially zero.
\item The effect of the regridding procedure is exemplified in Fig. \ref{fig:constraint-diff-amps} where we plot a constraint over time along $\scri^+$ for simulations with different wave amplitudes~$a$. Although the constraints begin increasing when regridding starts, they do not become unreasonable before getting very close to $i^+$. Fig. \ref{fig:constraint-diff-amps} also shows a convergence test of a constraint at a fixed angle across the entire radial domain.
\item The asymptotic Einstein condition $\nabla_{A'(A)}\nabla_{B)B'}\Theta=0$, which is valid on $\scri^+$, gives rise to the conformally invariant relations $\eth_c A = 0 = \eth_c' A$ and $\thorn_c'A = 0$, in addition to the equations $\sigma'=0=\kappa'$ with respect to the adapted frame. All these equations  are satisfied.
\item A variety of the spin-coefficient equations (Eq. (4.12.32) of \cite{Penrose:1984a}) were shown to be satisfied with respect to the frame on $\scri^+$ as well as across the timeslices. These contain spin-coefficients, $\eth,\eth',\thorn,\thorn'$ acting on spin-coefficients, $\Lambda$ and components of $\Psi_{ABCD}$ and $\Phi_{ABA'B'}$. These confirmed the interplay between the above quantities and operators.
\item A variety of the Bianchi identity equations (Eq. (4.12.36-40) of \cite{Penrose:1984a}) were shown to be satisfied on $\scri^+$ as well as across the timeslices. These contain spin-coefficients, $\Lambda$ and components of $\Psi_{ABCD}$ and $\Phi_{ABA'B'}$ and their $\eth,\eth',\thorn,\thorn'$ derivatives. This further showed the interplay with the above quantities and derivative operators, as well as the relation $A\psi_i = -\thorn\Psi_i$.
\item The Gauß-Bonnet theorem on a cut $C$ of $\scri^+$ is written $\int_{C}k\textrm{d}^2\Sigma = \int_{S^2}kJ\textrm{d}^2S = 4\pi$. This can be evaluated numerically and was shown to be satisfied along $\scri^+$. This is used as a confirmation for the correctness of the expressions for the area-element $J$ and the Gauß curvature $k$.
\item The reality of the Bondi energy (i.e., the vanishing of the imaginary part of the integral) is satisfied, which requires a non-trivial interplay between the imaginary parts of several quantities that appear in the mass-aspect.
\item There are two representations of the mass-aspect used in the definition of the Bondi components \cite{Frauendiener:2021a}. These are made up of completely different system variables and the Bondi energy resulting from using these two different mass aspects has been found to agree up to numerical precision.
\item The Bondi-Sachs mass-loss given by Eq.~\eref{eq:BSML} is satisfied. This is shown in Fig.~\ref{fig:BEComparison}.
\end{itemize}

\begin{figure}[H]
    \centering
    \subfloat[\centering ]
    {{\includegraphics[width=0.5\linewidth]{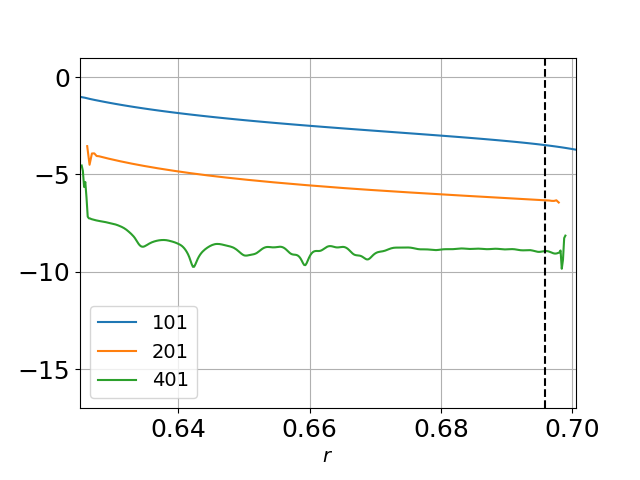}}}
    \qquad
    \subfloat[\centering ]
    {{\includegraphics[width=0.5\linewidth]{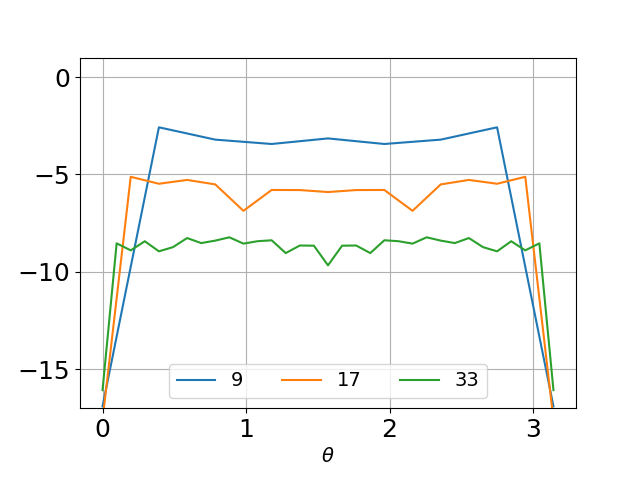}}}
    \caption{A convergence test of the $\log_{10}$ of the modulus of a component of Bianchi constraint $G_{AB}$ at $t = 1.65$ at: (a) a fixed angle and (b) a fixed radius close to $\mathscr{I}^+$ in the situation described in Sec.~\ref{sec:general-situation}. In (b) the vertical dashed line represents the location of $\scri^+$. The curves representing coarse to fine resolution are given from top to bottom.}
    \label{fig:convergence_tests}
\end{figure}

\subsection{A global evolution and ringing}\label{sec:general-situation}
In this section we show visualizations of the space-time resulting from the setup described in Sec.~\ref{sec:IDandBD} using $a=1$ in the boundary condition for $\psi_0$, to get an idea of the main features. 

\begin{figure}[H]
    \centering
    \subfloat[\centering $|(\Psi_0)_2|$]
    {{\includegraphics[width=0.5\linewidth]{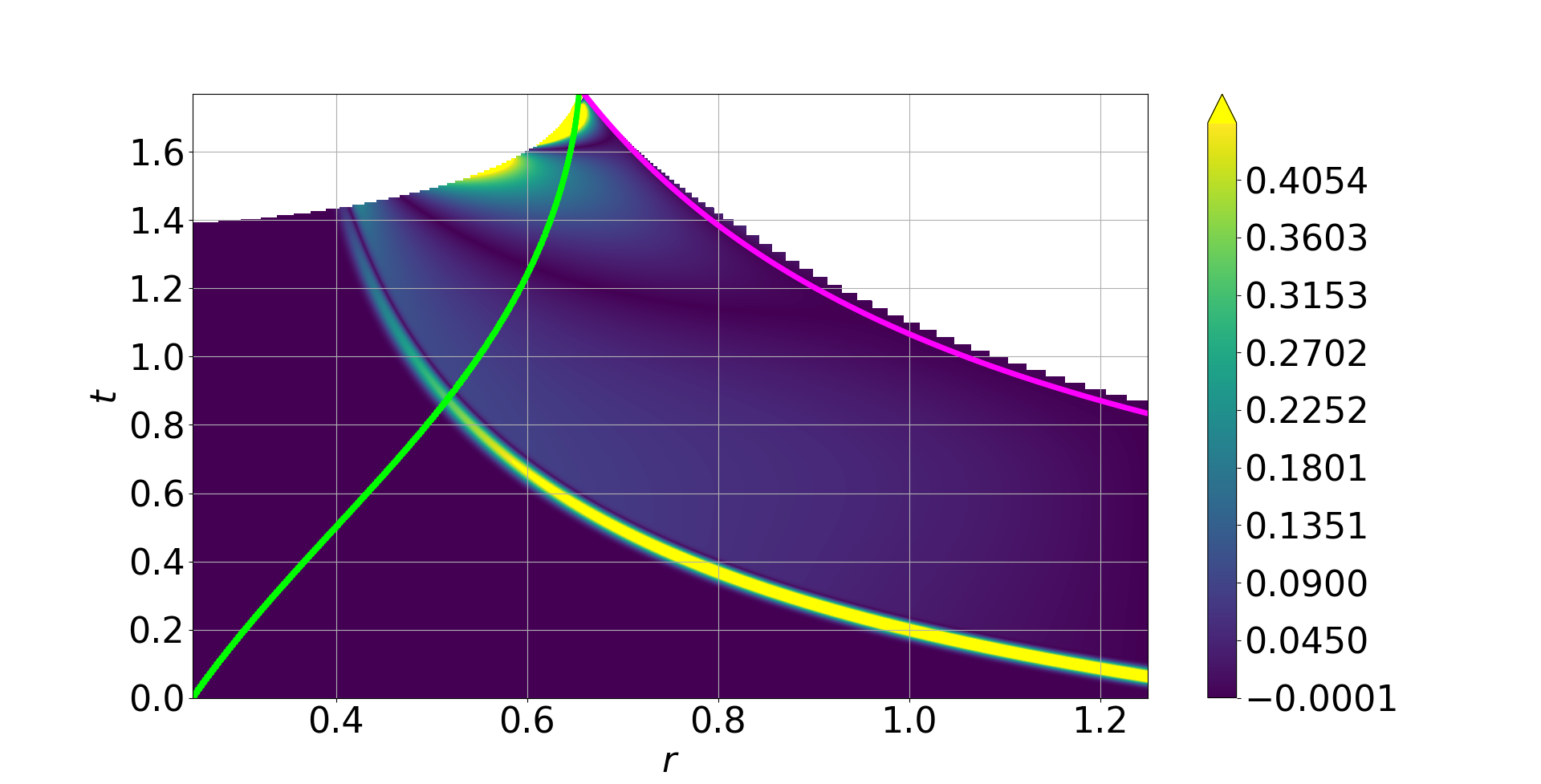}}}
    \qquad
    \subfloat[\centering $|(\Psi_0)_2|$ zoomed in.]
    {{\includegraphics[width=0.5\linewidth]{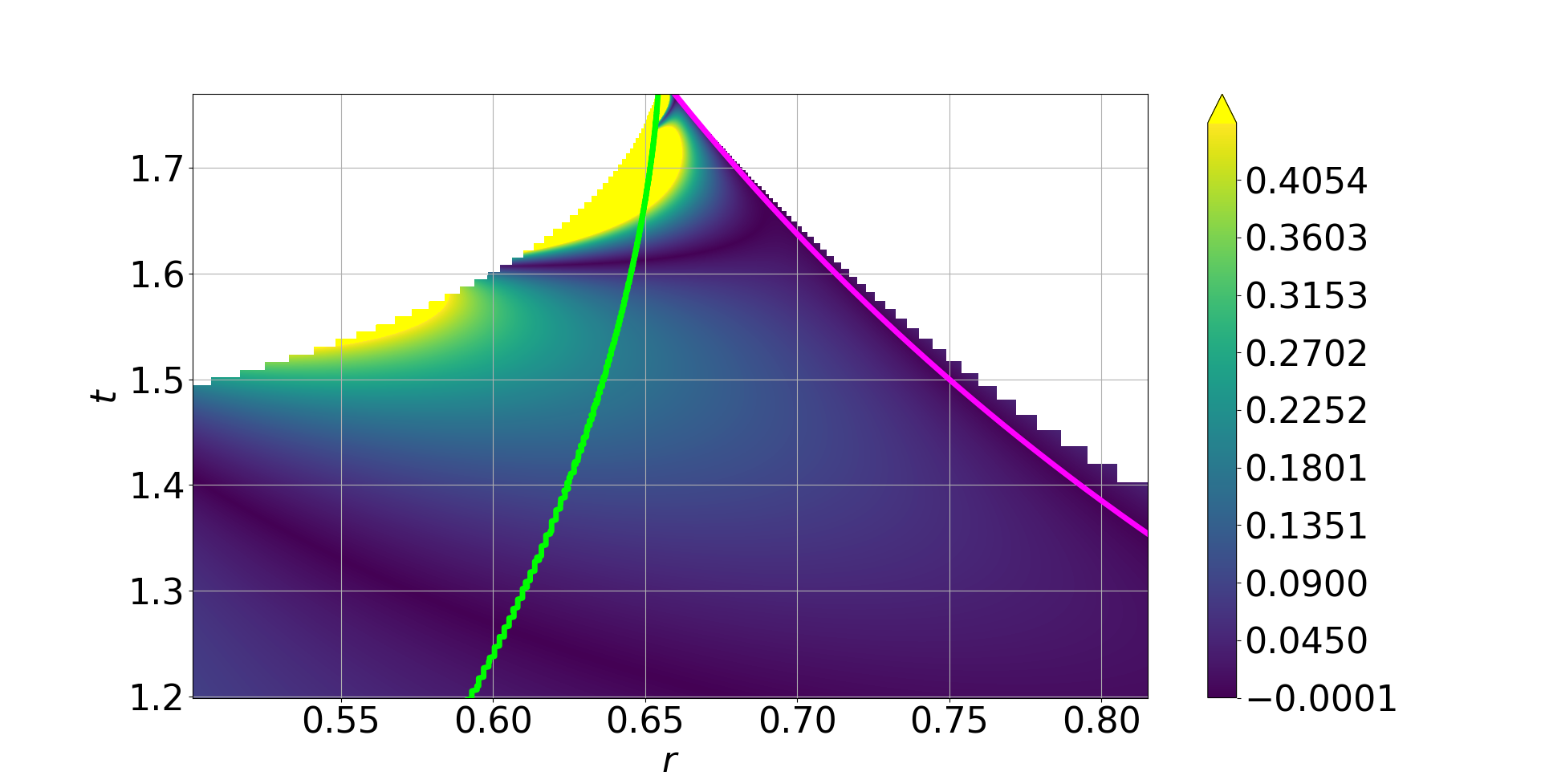}}}
	\\
	\subfloat[\centering $|(\Psi_4)_2|$]
    {{\includegraphics[width=0.5\linewidth]{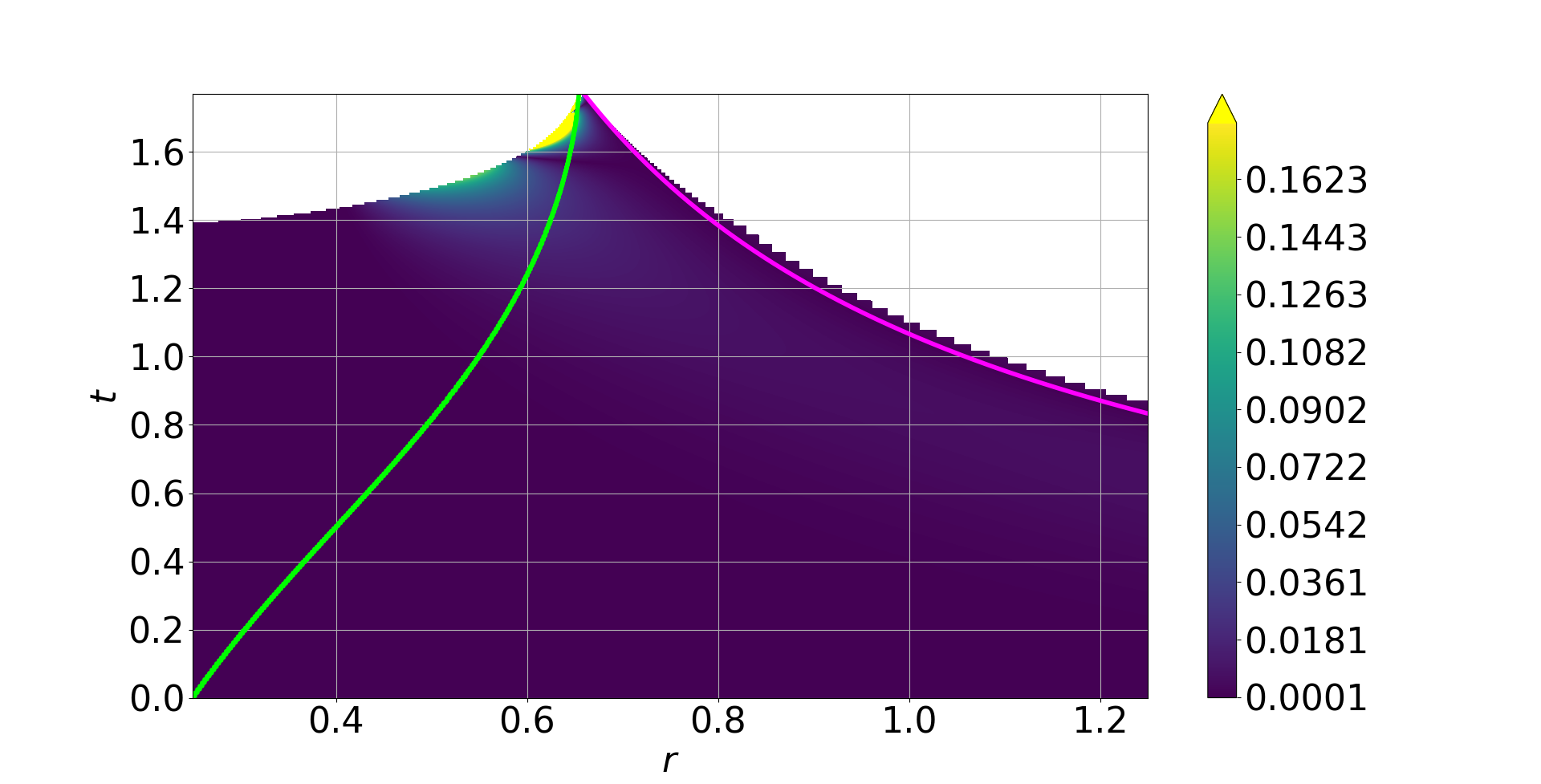}}}
    \qquad
    \subfloat[\centering $|(\Psi_4)_2|$ zoomed in.]
    {{\includegraphics[width=0.5\linewidth]{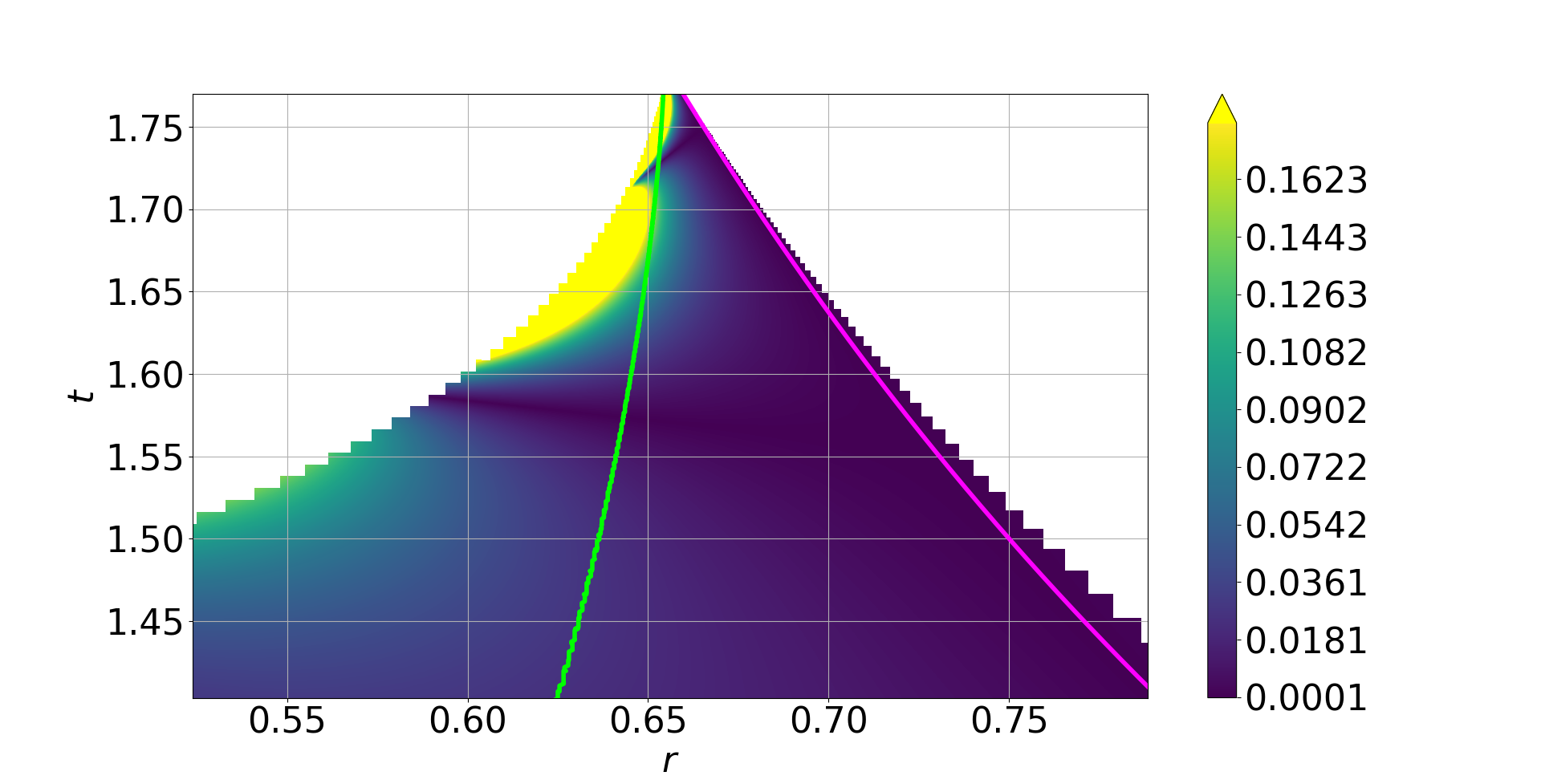}}}
    \caption{Contour plots of the modulus of the $l=2$ mode of the Weyl spinor components $\Psi_0$ and $\Psi_4$ over the space-time with the initial wave amplitude $a=1$. Note that the levels for (a) were chosen to resolve the detail in the ``ringing'' rather than the initial part of the ingoing wave.}
    \label{fig:ContourPlotspsi0psi4}
\end{figure}

Fig.~\ref{fig:ContourPlotspsi0psi4}(a) shows a contour plot of $\Psi_0$ over the entire computed space-time for a fixed angle. Note, that this is the Weyl spinor component, which vanishes on $\scri$  and not the gravitational spinor. The curve emanating from the bottom left corner is our approximation to the MOTS, as detailed in Sec.~\ref{sec:MOTS}. One can see the path of the ingoing wave emanating from just above the bottom right corner. Our regridding procedure results in the continual trimming of the spatial computational domain, starting on the left boundary just before $t =1.4$ and just after $t =0.8$ on the right boundary. As expected, the components of $\Psi_{ABCD}$ and hence $\psi_{ABCD}$ diverge as we approach $i^+$, which has a slightly different location to exact Schwarzschild.

Importantly, Fig.~\ref{fig:ContourPlotspsi0psi4} also showcases a periodic behaviour in $\Psi_0$ and $\Psi_4$, most notably close to the left boundary, but also appearing outside of the black hole. It appears there are three periods before the space-time becomes extremely compactified close to $i^+$. This seems to be an effect associated with the expected ``ringing'' of a perturbed black hole. This behaviour which can be described in terms of \emph{quasinormal modes}, has been the subject of extensive studies (see \cite{Kokkotas:1999} for a review). However, this is linear theory and there are several differences to what we seem to see.

First, the quasinormal modes are determined outside the horizon of the unperturbed black hole. In fact, they are fixed by the boundary condition of vanishing excitation at the horizon. In the diagram it is obvious that the excitation of the black hole penetrates into the horizon and seems to come arbitrarily close to the singularity. Thus, there is no immediate connection between our non-linear excitation and the quasinormal modes even though this might be the case near $\scri$.

Second, we can also see that our approximation of the event horizon by an approximate MOTS is increasingly bad towards time-like infinity since we would expect that the event horizon, $\scri$ and the singularity are all approaching the same ``point'' as they do in the exact Schwarzschild solution. This means that we should not be too alarmed by the fact that in the diagram $(d)$ it looks like there is information coming out of the horizon. This is almost certainly not true since the event horizon will be located to the right of the green line. 

Third, we seem to see only three ``rings''. This might be related to the use of the conformal Gauß gauge. A significant feature of this gauge is the relationship between the proper time $\tau$ and conformal time $t$ which is given along a spatially constant curve by the equation $\textrm{d}\tau^2 = \Theta^{-2}\textrm{d}t^2$. As the conformal factor $\Theta$ is known analytically \emph{a priori}, we can explicitly find the relationship along a curve with constant $r=r_c$
\begin{equation}\label{eq:ProperPhysicalTime}
	\tau = \frac{\Big{(}\frac{m}{2} + r_c\Big{)}^3}{r_c\Big{(}\frac{m}{2} - r_c\Big{)}}\textrm{arctanh}\Big{[}\frac{t}{r_c}\Big{(}\Big{(}\frac{m}{2}\Big{)}^2 - r_c^2\Big{)}\Big{]}.
\end{equation}
Depending upon the study in question, this is potentially problematic as the $\mathrm{arctanh}$ function maps the entire real line onto the interval $(-1,1)$. In fact, along $r_{i^+} = (3 + \sqrt{5})m/4$ at $t \approx1.75$ we have $\tau \approx 6.31$. This implies that the remaining space-time is roughly compactified into the conformal time interval $[1.75,\,t_{i^+}\approx1.789]$. For a study of the quasinormal modes, this compactification will hide important features as the conformal time-step during the numerical evolution will ultimately step over the ringing. It is guaranteed that at some point the conformal time-step required to accurately resolve the physical ringing will be less than machine-precision. However, any compactification in time, will result in the last time-step in conformal time covering an infinite amount of physical time. Thus, there is a trade-off between trying to determine as many ``rings'' as possible and approaching time-like infinity as closely as possible.

There might be a possibility to increase the number of periods that we can detect. This is due to a special property of conformal geodesics. In contrast to affine geodesics for which an affine parameter is fixed up to an affine transformation $t\mapsto at+b$, the conformal parameters can be reparameterized by M\"{o}bius transformations in $t$
\begin{equation}
	t  \mapsto \frac{a t  + b}{c t  + d},
\end{equation}
which could possibly be helpful to allow a more detailed look at this ringing. Choosing suitable transformations may delay the arrival at $i^+$ and therefore provide a more detailed look at the ringing. However, this discussion is outside the scope of this paper and will be left for future work.

\subsection{The Bondi-Sachs energy and mass-loss}
\label{sec:BSML}

\begin{figure}[H]
    \centering
    \subfloat[\centering Bondi energy]
    {{\includegraphics[width=0.5\linewidth]{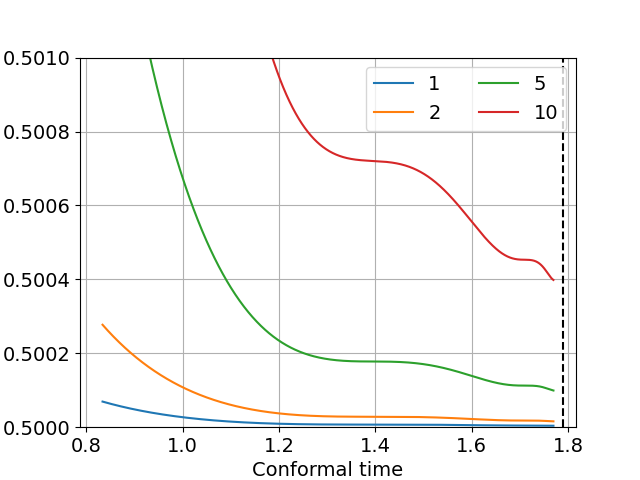}}}
    \qquad
    \subfloat[\centering Error in the Bondi-Sachs mass-loss]
    {{\includegraphics[width=0.5\linewidth]{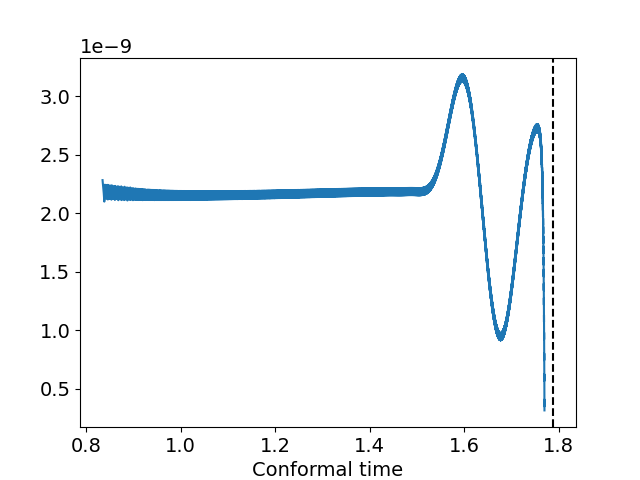}}}
    \caption{(a) The Bondi energy for different initial wave amplitudes $a$ and (b) the error in the satisfaction of the Bondi-Sachs mass-loss for $a=10$ along $\mathscr{I}^+$.}
    \label{fig:BEComparison}
\end{figure}

We performed five simulations with the amplitude parameter $a$ in the wave profile fixed as $1,2,5$ and $10$. They evolved up to $t=1.77$ at which point the system variables start to diverge beyond what the numerics can handle due to the close proximity of $i^+$. Fig.~\ref{fig:BEComparison} (a) shows how the Bondi energy changes with the amplitude of the impinging wave. This shows a ``wiggling'' which is probably associated with the ringing as seen in Fig.~\ref{fig:ContourPlotspsi0psi4}. It is clear that as $a\rightarrow0$ the Bondi energy decreases down towards the original Schwarzschild mass $m=0.5$. Fig.~\ref{fig:BEComparison} (b) shows the Bondi-Sachs mass-loss formula given by Eq.~\eref{eq:BSML} is satisfied to very good precision. It also shows an oscillatory behaviour (which is still of the order 1e-9) beginning at around $t=1.5$. This is roughly when the periodic behaviour begins in Fig.~\ref{fig:ContourPlotspsi0psi4} and the wiggling starts in Fig.~\ref{fig:BEComparison} (a). It is then feasible to conclude that this is the time when we transition from pure backreaction to the black hole's "ringing". Table~\ref{tab:BEs} presents the difference of the Bondi energies on the first cut of $\mathscr{I}^+$ and the initial mass $m=\frac12$ of the Schwarzschild black hole for different wave amplitudes $a$ (to 5 decimal places). It is clearly seen that this scales like $a^2$, i.e., the energy of the space-time due to the ingoing gravitational wave is quadratic in its amplitude, as one would expect from the linearized theory of gravitational waves.

\begin{table}[!h]
	\begin{center}
	\begin{tabular}{ |l|l|l|l|l| } 
	 \hline
	 $a$ & 1 & 2 & 5 & 10 \\ 
	 \hline
	 $m_0[\Omega]-m$ & 0.00007 & 0.00028 & 0.00173 & 0.00685 \\ 
	 \hline
	\end{tabular}
	\caption{The Bondi energy evaluated on the first cut of $\mathscr{I}^+$ for different initial wave amplitudes $a$.}
	\label{tab:BEs}
	\end{center}
\end{table}

Although there is still an infinite amount of physical time between the end of our simulation at $t =1.77$ and where $\scri^+$ meets $i^+$, Fig.~\ref{fig:BEComparison} suggests the Bondi energy might be settling down to values different to the original Schwarzschild mass $m$. This would imply that the wave imparted more energy to the black hole than was lost through the ringing response. 
To answer this question in more detail it is necessary to extend the evolution time.

\section{Summary and discussion}
\label{sec:Summary}

In this paper we presented a framework for studying the interaction between gravitational waves hitting a black hole. This is achieved by setting up and solving an initial boundary value problem for the general conformal field equations. We described the effect of a gravitational wave impinging on an initially spherically symmetric (Schwarzschild) black hole. The advantage of using the GCFE together with the conformal Gauß gauge is that the time-like boundary eventually reaches and intersects null infinity so that the full information of the outgoing radiation becomes available. This allows us to evaluate the expressions for the Bondi-Sachs energy-momentum and numerically verify the validity of the mass loss formula directly on $\scri$.

After the gravitational wave is sent into the space-time we observe a ``periodic'' behaviour in the components of the Weyl spinor (and other field quantities as well). It is tempting to relate these to the quasinormal modes of a linearly perturbed Schwarzschild black hole which have been extensively studied previously. However, this is not straightforward since those satisfy different boundary conditions from what we find in our simulation. While the linear modes are forced to vanish on the horizon, our excitation definitely persist on and even inside the horizon. This raises the question as to whether there is a difference in the characteristic frequencies between the linear and our non-linear modes and how would one find out?

The quasinormal modes are commonly computed by solving the radial Regge-Wheeler-Zerilli equation \cite{Regge:1957,Zerilli:1970} for a master function representing a linear perturbation of the Schwarzschild metric. This is then presented in Schwarzschild time along curves of constant Schwarzschild radius. However, in our situation we have neither, as our non-linear gravitational perturbations couple to the geometry, causing a physical deviation from the Schwarzschild metric. This makes it difficult to make physically meaningful comparisons.

Due to the use of the conformal Gauß gauge which has the property that the relationship between the physical (proper or retarded) times and the conformal time parameter is ultimately exponential we see the ``ringing'' currently only for three periods. This is definitely not enough to make any long-term comparisons with the linear regime. However, we pointed out that rescaling the conformal parameter could be used to extend the simulation, in principle indefinitely. Another possibility would be to abolish the conformal Gauß gauge altogether and use a different choice. However, this would entail non-trivial changes to the system of equations. In particular, we would lose the simplicity of the GCFE with the clear splitting into wave-type equations for the Weyl curvature and advection equations for the remaining variables. This would have the consequence that the IBVP would become much more complicated and one would have to discuss much more complicated boundary conditions.

The framework as presented here seems to be ideal to study the interaction of black holes and gravitational waves in a very clean setting. The possibility of setting up initially unperturbed situations and injecting the perturbation from the boundary opens the way to study many more questions of principle. For example, our next task will be to study whether we can ``kick'' a black hole by injecting gravitational waves with a linear momentum into the space-time. What would the response be? It should be straightforward also to replace the initial Schwarzschild black hole with a Kerr or Reissner-Nordström black hole. This would allow us to study phenomena such as super-radiance, the effect of gravitational perturbations on the inner Cauchy horizons and the stability of the solutions in general.

\ack

CS thanks Nigel Bishop and Florian Beyer for helpful discussions. This work was supported by the Marsden Fund Council from Government funding, managed by Royal Society Te Apārangi.

\appendix
\section{The constraint equations}
\label{sec:constraint-equations}

\begin{subequations}\label{constreqs}

The constraint equations for the GCFE in the conformal Gauß gauge read as follows  
\begin{align}
	0 = Z^0_{AB} &:= \del^C{}_{(A}c^0_{B)C} + \frac{1}{\sqrt{2}}K_{(A}{}^C{}_{B)C},\label{constr:6}  \\
	0 = Z^i_{AB} &:= \del^C{}_{(A}c^i_{B)C}, \qquad i=1,2,3.\label{constr:7}\\
  0 = J_{ABC}  &:= -\del_{(A}{}^E\gamma_{B)EC} + \frac14f_{AB}f_{CD}o^D + \frac12K_{(A}{}^E{}_{|C|}{}^DK_{B)EFD}o^F \nonumber\\
  &- \frac18\left( P_{AC(BD)} + P_{BC(AD)} \right)o^D - \frac18\left( P_{AD(BC)} + P_{BD(AC)} \right)o^D \nonumber \\
  &+ \frac14\left( P_A{}^D{}_{BD} + P_B{}^D{}_{AD} \right)o_C 
		+ \frac18\left( P_A{}^D{}_{(CD)}o_B + P_B{}^D{}_{(CD)}o_A \right) \label{constr:5} \\
	&- \frac18\left( \eps_{AC}P_B{}^E{}_{(DE)} + \eps_{BC}P_A{}^E{}_{(DE)} \right)o^D - 
		\frac12o^D\del_{D(A}f_{B)C} + \frac12o_{(A}\del_{B)}{}^Ef_{CE}, \nonumber \\
  0 = Z_{ABCD} &:= -\del_{(A}{}^EK_{B)ECD} - \frac12f_{CD}K_{(A}{}^E{}_{B)E} - \frac12f_{C(A}K_{B)}{}^E{}_{DE} - \frac12\eps_{C(A}f^{EF}K_{B)DEF} \nonumber \\
                     &- \frac12\eps_{D(A}f^{EF}K_{B)ECF} + \frac12\eps_{C(A}P_{B)DE}{}^E + \frac12\eps_{D(A}P_{B)CE}{}^E \nonumber\\
  & + \frac12\Theta\psi_{ABCD} - \frac12\Theta\hat\psi_{ABCD},\label{constr:3} \\
	0 = T_{AB}   &:= \del_{(A}{}^Ef_{B)E} + \frac12P_{(A}{}^E{}_{B)E} - \frac12P_{E(A}{}^E{}_{B)}, \label{constr:4}\\
  0 = U_{ABCD} &:= -\del_{(A}{}^EP_{B)ECD} - \frac12f_{CE}P_{(A}{}^E{}_{B)D} - \frac12f_{DF}P_{(A}{}^E{}_{|C|B)} \nonumber\\
  &-  \frac12\left( f_{D(A}P_{B)}{}^E{}_{CE} + f_{C(A}P_{B)}{}^E{}_{ED} \right) 
  + \frac12P_{(A}{}^E{}_{|C|}{}^FK_{B)EDF} - \frac12P_{(A}{}^{EF}{}_{|D|}K_{B)ECF} \nonumber\\
  &+ \frac12\psi_{ABCE}h^E{}_D + \frac12\hat\psi_{ABDE}h_C{}^E, \label{constr:1}\\
	0 = G_{AB} &:= \del^{CD}\psi_{ABCD} + K^{CE}{}_E{}^D\psi_{ABCD} + K^{CDE}{}_{(A}\psi_{B)CDE} \label{constr:2}.
\end{align}
\end{subequations}

\printbibliography
\end{document}